\documentclass[prl,twocolumn,superscriptaddress]{revtex4}

\newcommand{\FigDirectory}{.}

\usepackage{amsmath}                          
\usepackage{amsfonts}                         
\usepackage{amssymb}                          

\usepackage{dsfont}                           

\usepackage{float}                            
\usepackage{graphicx}                         
\usepackage{subfigure}			              

\usepackage[colorlinks=true,
            urlcolor=blue,
            citecolor=blue,
            linkcolor=black,
            pdfstartview=FitH,
            pdfpagemode=None]{hyperref}

\newcommand{\lsim}{\mathrel{\raise.3ex\hbox{$<$\kern-.75em\lower1ex\hbox{$\sim$}}}}
\newcommand{\gsim}{\mathrel{\raise.3ex\hbox{$>$\kern-.75em\lower1ex\hbox{$\sim$}}}}

\def\QECCnk[[#1,#2]]{[\![#1, #2]\!]}

\def\QECCnkq[[#1,#2,#3]]{[\![#1, #2]\!]_{#3}^{\vphantom{T}}}

\def\QECCnkd[[#1,#2,#3]]{[\![#1, #2, #3]\!]}

\def\QECCnkdq[[#1,#2,#3,#4]]{[\![#1, #2, #3]\!]_{#4}^{\vphantom{T}}}

\def\QECCnkgd[[#1,#2,#3,#4]]{[\![#1, #2, #3, #4]\!]}

\def\QECCnkgdq[[#1,#2,#3,#4,#5]]{[\![#1, #2, #3, #4]\!]_{#5}^{\vphantom{T}}}

\def\QECCnkdc[[#1,#2,#3,#4]]{[\![#1, #2, #3; #4]\!]}

\def\QECCnkdcq[[#1,#2,#3,#4,#5]]{[\![#1, #2, #3; #4]\!]_{#5}^{\vphantom{T}}}

\def\QECCnkgdcq[[#1,#2,#3,#4,#5,#6]]{%
  [\![#1, #2, #3, #4; #5]\!]_{#6}^{\vphantom{T}}}

\def\openone{\leavevmode\hbox{\small1\normalsize\kern-.33em1}}

\long\def\symbolfootnote[#1]#2{\begingroup%
\def\thefootnote{\fnsymbol{footnote}}\footnote[#1]{#2}\endgroup}

\begin{document}


%
\title{The surface code on the rhombic dodecahedron}

\author{Andrew J. \surname{Landahl}}
\email[]{alandahl@sandia.gov}
\affiliation{Center for Computing Research,
             Sandia National Laboratories,
             Albuquerque, NM, 87185, USA}
\affiliation{Center for Quantum Information and Control,
             University of New Mexico,
             Albuquerque, NM, 87131, USA}
\affiliation{Department of Physics and Astronomy,
             University of New Mexico,
             Albuquerque, NM, 87131, USA}


\begin{abstract}

I present a jaunty little $[[14, 3, 3]]$ non-CSS surface code that can be
described using a rhombic dodecahedron.  Do with it what you will.

\end{abstract}

%
\maketitle


%

\begin{figure}[!ht]
\center{
\subfigure[\ 3D wireframe.]{%
\includegraphics[width = 0.33\columnwidth]{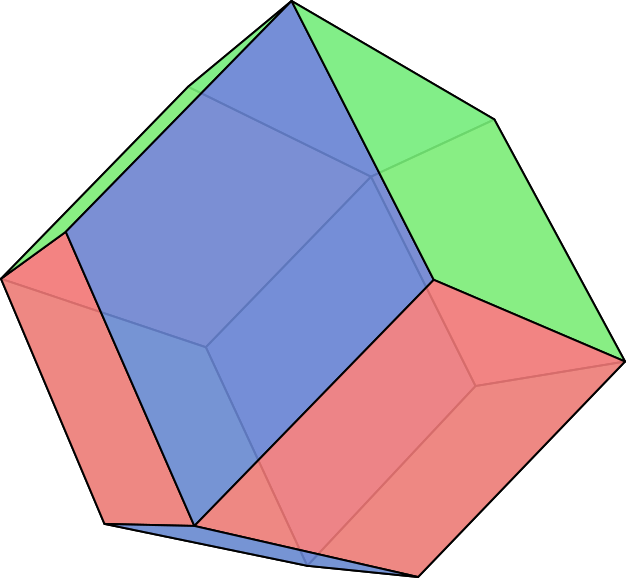}
\label{fig:3d-rdpic}
} 
\hspace{1em}
\subfigure[\ Azimuthal projection.]{%
\includegraphics[width = 0.33\columnwidth]{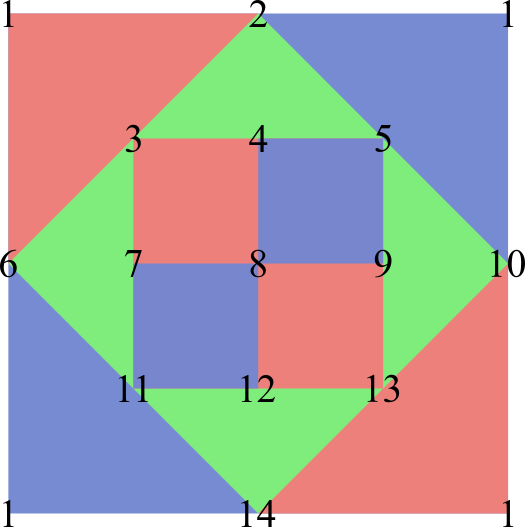}
\label{fig:fisheye-rdpic}
} 
\hspace{1em}
\subfigure[\ 2D net.]{%
\includegraphics[width = 0.75\columnwidth]{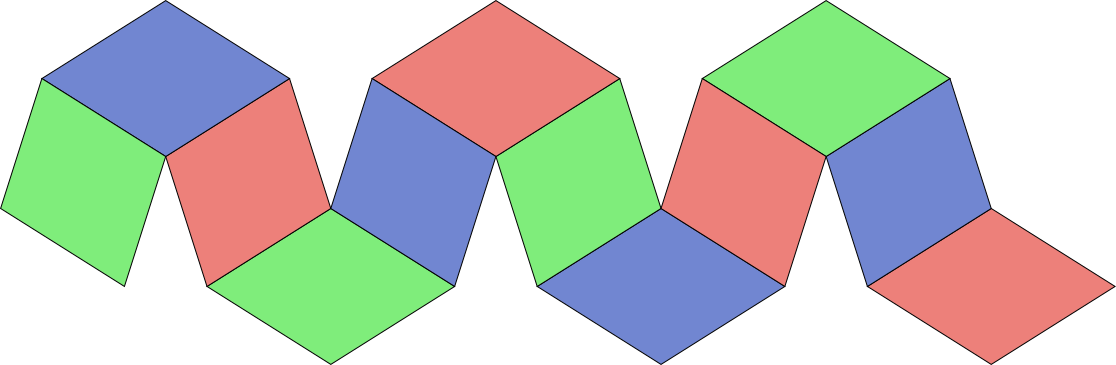}
} 
} 
\caption{\label{fig:rhombic-dodecahedron}(Color online.) Depictions of the 
rhombic dodecahedron.}
\end{figure}

\noindent \textit{The surface.}
Figure~\ref{fig:rhombic-dodecahedron} depicts the rhombic dodecahedron in
wireframe, azimuthal, and net projections.  Enlarge, cut out, fold, and tape
the net diagram together to create your own handy three-dimensional model.
Better yet, print one using a 3D printer for maximum
enjoyment.\endnote{Thanks to Matt Curry and Jaimie Stephens for 3D printing
a rhombic dodecahedron for me.}

The rhombic dodecahedron has 14 vertices, 24 edges, and 12 faces.  Eight
vertices have degree three and six vertices have degree four.  Each of its
dozen faces is a rhombus of the same size and shape, with the long diagonal
of each rhombus having a length that is $\sqrt{2}$ times the length of the
short diagonal.  (The Bilinski rhombic dodecahedron has a slightly different
geometry, for golden ratio aficionados \cite{Bilinski:1960a,
Gruenbaum:2010a}.) The faces can be colored red, green, and blue so that no two
adjacent faces share the same color.




%

\noindent \textit{The code.}
Locate a qubit on each vertex and a check on each face of the rhombic
dodecahedron; $X$, $Y$, and $Z$ Pauli checks are on the
the red, green, and blue faces respectively.  Because the product of all
face checks is the identity, there are only 11 independent checks, meaning
that the code encodes three logical qubits.  Another way to arrive at this
conclusion is to interpret the degree-three vertices as disclination
twists~\footnote{The word ``twist'' is used ambiguously in the literature on
lattice defects in topological codes~\cite{Kitaev:2012a} to mean either a
lateral dislocation~\cite{Bombin:2010b} or a rotational
disclination~\cite{Yoder:2017a}.  I use the word
``disclination'' to clarify which I mean.}, where each twist after the
first pair contributes a quantum dimension of $\sqrt{2}$ to the logical
Hilbert space, following the quasiparticle-based reasoning discussed in
Ref.~\cite{Brown:2017a}.

Subject to the vertex labeling and face coloring in the figures, the
algebraic representations of the checks are the $S_i^j$ operators below,
with the operator $S_4^Z$ written in parentheses because it is not an
independent check---it is the product of the preceding ones.
\begin{align}
\setlength\arraycolsep{0.05pt}
\begin{array}{*{16}c}
\phantom{S_1^X} &\phantom{\ =} &\phantom{X} &\phantom{X} &\phantom{X}
&\phantom{X} &\phantom{X} &\phantom{X} &\phantom{X} &\phantom{X}
&\phantom{X} &\phantom{X} &\phantom{X} &\phantom{X} &\phantom{X}
&\phantom{X}
\\
S_1^X &\ = &\ X & X & X & I & I & X & I & I & I & I & I & I & I & I \\[1ex]
S_2^X &\ = &\ I & I & X & X & I & I & X & X & I & I & I & I & I & I \\[1ex]
S_3^X &\ = &\ I & I & I & I & I & I & I & X & X & I & I & X & X & I \\[1ex]
S_4^X &\ = &\ X & I & I & I & I & I & I & I & I & X & I & I & X & X \\[1ex]
S_1^Y &\ = &\ I & Y & Y & Y & Y & I & I & I & I & I & I & I & I & I \\[1ex]
S_2^Y &\ = &\ I & I & Y & I & I & Y & Y & I & I & I & I & Y & I & I \\[1ex]
S_3^Y &\ = &\ I & I & I & I & Y & I & I & I & Y & Y & I & I & Y & I \\[1ex]
S_4^Y &\ = &\ I & I & I & I & I & I & I & I & I & I & Y & Y & Y & Y \\[1ex]
S_1^Z &\ = &\ Z & Z & I & I & Z & I & I & I & I & Z & I & I & I & I \\[1ex]
S_2^Z &\ = &\ I & I & I & Z & Z & I & I & Z & Z & I & I & I & I & I \\[1ex]
S_3^Z &\ = &\ I & I & I & I & I & I & Z & Z & I & I & Z & Z & I & I \\[1ex]
(S_4^Z &\ = &\ Z & I & I & I & I & Z & I & I & I & I & Z & I & I & Z).
\end{array}
\end{align}

It may seem puzzling at first that a surface code on an object that is
topologically equivalent to a sphere can hold
\emph{any} logical qubits because the surface code is a homological code,
meaning that the logical operators have a homological character, yet the
homology of a sphere is trivial.  The resolution to this puzzle is the
presence of the disclination twist defects---they provide locations on which
string-like logical operators can begin and end.

We can construct a basis for the code's logical space in the following way.
Each collection of faces of the same color forms an ``equator'' on the
rhombic dodecahedron.  Along each of the three colored equators, construct a
weight-four Pauli operator that commutes with all of the checks and other
equator operators but which cannot be expressed as a product of them---these
are representatives for the logical $\overline{X}$ operators for the code.
Algebraically, the operators are
\begin{align}
\setlength\arraycolsep{0.05pt}
\begin{array}{*{16}c}
\phantom{\overline{X}_g} &\phantom{\ =} &\phantom{X} &\phantom{X}
&\phantom{X} &\phantom{X} &\phantom{X} &\phantom{X} &\phantom{X}
&\phantom{X} &\phantom{X} &\phantom{X} &\phantom{X} &\phantom{X}
&\phantom{X} &\phantom{X} 
\\
\overline{X}_r &\ =
&\ Z & I & Y & I & I & I & I & Z & I & I & I & I & Y & I \\[1ex]
\overline{X}_g &\ =
&\ I & I & X & I & Z & I & I & I & I & I & Z & I & X & I \\[1ex]
\overline{X}_b &\ =
 &\ X & I & I & I & Y & I & I & X & I & I & Y & I & I & I.
\end{array}
\end{align}

To each logical $\overline{X}$ operator, associate a logical $\overline{Z}$
operator in the following way.  For each ``equator,'' locate a collection of
three qubits that forms a ``V'' shape that points along the equator; the
``V'' is supported on a string of qubits that connects twist defects on
either side of the equator.  Choose these three ``V'' sets so that they
correspond to disjoint sets of qubits.  On each of these ``V'' sets, form a
Pauli operator that anticommutes with the corresponding logical
$\overline{X}$ operator on the equator but commutes with all other checks
and logical operators.  Algebraically, one choice for the logical
$\overline{Z}$ operators is
\begin{align}
\label{eq:logical-z-ops}
\setlength\arraycolsep{0.05pt}
\begin{array}{*{16}c}
\phantom{\overline{X}_g} &\phantom{\ =} &\phantom{X} &\phantom{X}
&\phantom{X} &\phantom{X} &\phantom{X} &\phantom{X} &\phantom{X}
&\phantom{X} &\phantom{X} &\phantom{X} &\phantom{X} &\phantom{X}
&\phantom{X} &\phantom{X} 
\\
\overline{Z}_r &\ =
&\ I & I & X & Z & I & I & Z & I & I & I & I & I & I & I  \\[1ex]
\overline{Z}_g &\ =
&\ I & I & I & I & Y & I & I & I & X & X & I & I & I & I \\[1ex]
\overline{Z}_b &\ =
&\ Z & I & I & I & I & Y & I & I & I & I & I & I & I & Y.
\end{array}
\end{align}

\begin{figure}[!ht]
\center{
\subfigure[\ Logical $X$ operators.]{%
\includegraphics[width = 0.33\columnwidth]{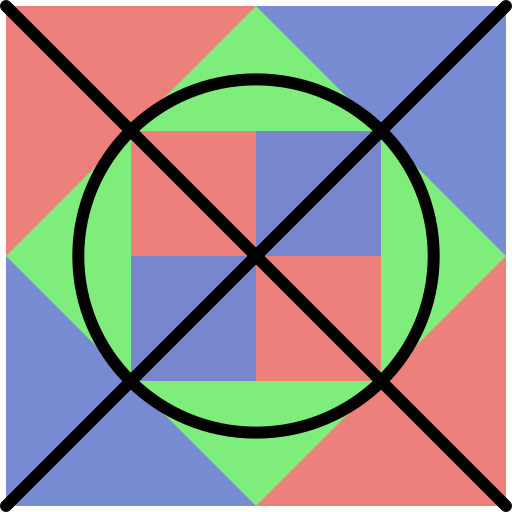}
\label{fig:rd-x}
} 
\hspace{1em}
\subfigure[\ Logical $Z$ operators.]{%
\includegraphics[width = 0.33\columnwidth]{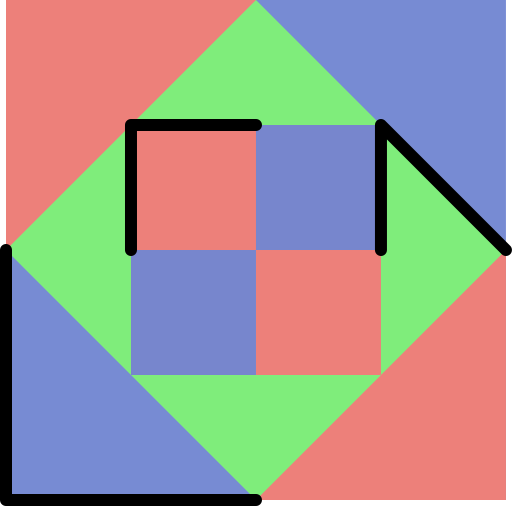}
\label{fig:rd-z}
} 
} 
\caption{\label{fig:logical-ops}(Color online.) A basis of logical operators
for the code.}
\end{figure}

Figure~\ref{fig:logical-ops} depicts the logical $X$ and $Z$ operators for
the code using the azimuthal projection of the rhombic dodecahedron.

Observe, from Eq.~(\ref{eq:logical-z-ops}), that the distance of the surface
code on the rhombic dodecahedron is at most three.  Using the symmetry of
the rhombic dodecahedron, it is a straightforward task to verify that all
weight-two Pauli operators anticommute with some check.  This proves that
three is, in fact, the minimum distance of the code.

Although standard surface codes \cite{Kitaev:1996a, Kitaev:1997a,
Kitaev:1997b, Dennis:2002a} are CSS codes \cite{Calderbank:1996a,
Steane:1996c}, the presence of disclination twist defects allows the
presence of $Y$ checks \cite{Yoder:2017a}, making it a non-CSS code.  The
consequence of all of these considerations is that the surface code on the
rhombic dodecahedron is a $[\![14, 3, 3]\!]$ non-CSS code.

%

\begin{figure}[!ht]

\center{
   \includegraphics[width = 0.33\columnwidth]{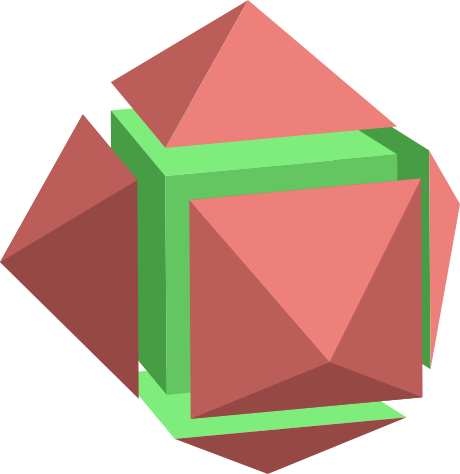}
}

\caption{\label{fig:cube-rd}(Color online.) The rhombic dodecahedron
can be formed by gluing pyramids to each face of the cube.}
\end{figure}

\noindent \textit{Ruminations.}
A natural follow-on question to ask is, ``Does the surface code on the
rhombic dodecahedron generalize to larger polyhedral codes?'' Perhaps it
does, but as far as I can tell, there is no unique systematic method of
generalization; there are an infinite number of larger face-three-colorable
polyhedra.  However, turning the other direction, the surface code on the
cube is an obvious progenitor of the surface code on the rhombic
dodecahedron.  (In fact, one can glue a pyramid to each face of the cube to
generate the rhombic dodecahedron%
, as depicted in Fig.~\ref{fig:cube-rd}%
.) This ``cubic surface code'' has eight qubits and eight disclination twist
defects, so three logical qubits total.  The distance turns out to be two,
so it is an $[\![8, 3, 2]\!]$ code, just like the (distinct!) ``smallest
interesting colour code'' \cite{Campbell:2016a}.  Perhaps you will be the
one to discover interesting generalizations of the rhombic docdecahedron
surface code.  Be sure to enjoy your time doing so.

%

%
\begin{acknowledgments}

\noindent \textit{Acknowledgments.}
Thanks to my many colleagues patient enough to listen to me muse about this
code while I whirled a 3D model of it around in my fingers, including (in
alphabetical order) Jonas Anderson, Dave Bacon, Steve Brierley, Ben Brown,
Bob Carr, Dan Browne, Stephen DiPippo, Anand Ganti, Tomas Jocyhm-O'Connor,
Cody Jones, Markus Kesselring, Isaac Kim, Vadym Kliuchnikov, Alex Kubica,
Setso Metodi, Naomi Nickerson, Tom O'Brien, John Preskill, Robert
Raussendorf, Ciar{\'{a}}n Ryan-Anderson, John Siirola, Jaimie Stephens,
Barbara Terhal, Dave Wecker, Nathan Wiebe, Wayne Witzel, and Ted Yoder.  A
special thanks to Jaimie Stephens for generating the 3D colored figures in
this paper.

This work was performed, in part, at the Aspen Center for Physics, which is
supported by National Science Foundation Grant PHY-1607611.
 
Sandia National Laboratories is a multimission laboratory managed and
operated by National Technology and Engineering Solutions of Sandia, LLC., a
wholly owned subsidiary of Honeywell International, Inc., for the U.S.
Department of Energy's National Nuclear Security Administration under
contract DE-NA-0003525.
 
This paper describes objective technical results and analysis. Any
subjective views or opinions that might be expressed in the paper do not
necessarily represent the views of the U.S. Department of Energy or the
United States Government.

\end{acknowledgments}

\bibliographystyle{landahl}
\bibliography{landahl}

\end{document}